\documentclass[pdflatex,sn-mathphys-num]{sn-jnl}
\usepackage{graphicx}%
\usepackage{multirow}%
\usepackage{amsmath,amssymb,amsfonts}%
\usepackage{amsthm}%
\usepackage{mathrsfs}%
\usepackage[title]{appendix}%
\usepackage{xcolor}%
\usepackage{textcomp}%
\usepackage{manyfoot}%
\usepackage{booktabs}%
\usepackage{algorithm}%
\usepackage{algorithmicx}%
\usepackage{algpseudocode}%
\usepackage{listings}%

\begin{document}

\title{Integrability of $R^2$ gravity cosmological models with radiation}

\author[1]{\fnm{Vsevolod~R.} \sur{Ivanov}}

\author[2]{\fnm{Sergey~Yu.} \sur{Vernov}}


\affil[1]{\orgdiv{Physics Department}, \orgname{Lomonosov Moscow State University}, \orgaddress{\street{Leninskie Gory 1}, \city{Moscow}, \postcode{119991}, \country{Russia}\email{vsvd.ivanov@gmail.com}}}

\affil[2]{\orgdiv{Skobeltsyn Institute of Nuclear Physics}, \orgname{Lomonosov Moscow State University}, \orgaddress{\street{Leninskie Gory 1}, \city{Moscow}, \postcode{119991}, \country{Russia}\email{svernov@theory.sinp.msu.ru}}}

\date{ / }

\abstract{We consider cosmological $R^2$ gravity models with radiation. We find the general solution to the trace equation $\Box R=0$ in the spatially flat Friedmann-Lemaitre-Robertson-Walker (FLRW) metric.
We analyze possible evolution of the Hubble parameter depending on the sign of the radiation energy density and find conditions for the existence of a bounce solution.
A scalar field Lagrangian with the induced gravity term and the fourth-order monomial potential can play a role of radiation. In this case, we also obtain the general solution to the field equation.
Therefore, the resulting $R^2$ gravity model with a scalar field is integrable in the spatially flat FLRW metric. Using a conformal metric transformation, we obtain a two-field chiral cosmological model that is also integrable in the spatially flat FLRW metric.}

\keywords{Chiral cosmological model, modified gravity, integrability}




\maketitle

\section{Introduction}

$F(R)$ gravity models are actively investigated~\cite{Sotiriou:2008rp,DeFelice:2010aj,Nojiri:2010wj,Capozziello:2011et,Nojiri:2017ncd}. The important feature of such models is the possibility to construct by the conformal metric transformation the corresponding model in the Einstein frame, in other words, the General Relativity model with one minimally coupled scalar field.  $F(R,\chi)$ models with the scalar field $\chi$ can be transformed to two-field models in the Einstein frame. Actions of the obtained two-field models have non-standard kinetic parts~\cite{Kaiser:2010ps}, in other words, they are chiral cosmological models (CCMs)~\cite{Chervon:1995jx,Starobinsky:2001xq,DiMarco:2002eb,Elizalde:2004mq,Chervon:2014dya,Kaneda:2015jma,Paliathanasis:2018vru,Christodoulidis:2019mkj,Chervon:2019nwq,Anguelova:2019omg,Giacomini:2020grc,Zhuravlev:2020ugb,Socorro:2020nsm,Fomin:2020caa,Fomin:2020woj,Paliathanasis:2020sfe,Braglia:2020fms,Braglia:2020eai,Anguelova:2020nzl,Paliathanasis:2021fxi,Ivanov:2024nnd,Pozdeeva:2025wsl}.

The most famous $F(R)$ gravity model is the Starobinsky $R^2$ inflationary model~\cite{Starobinsky:1980te} that includes both $R^2$ term and the Hilbert-Einstein term.
One of the most studied generalization of the Starobinsky model is the $R^2$-Higgs inflation with the induced gravity term and the fourth-order potential of the scalar field~\cite{Wang:2017fuy,Ema:2017rqn,Pi:2017gih,He:2018gyf,Ema:2019fdd,Gorbunov:2018llf,Bezrukov:2019ylq,He:2020ivk,Cheong:2019vzl,Gundhi:2020kzm,Gundhi:2020zvb,Kim:2025ikw}.

One of the motivations to consider the $F(R,\chi)$ gravity inflation models is the possibility to describe production of primordial black holes~\cite{Pi:2017gih,Cheong:2019vzl,Gundhi:2020kzm,Gundhi:2020zvb,Pozdeeva:2025wsl,Kim:2025dyi}. Note that the $R^2$ term arises as a quantum correction when inflationary models with scalar fields are considered~\cite{Steinwachs:2011zs,Salvio:2015kka,Elizalde:2015nya,Salvio:2021lka}.

The search for exact analytical solutions to evolution equations plays a significant role in the study of cosmological models. The majority of integrable cosmological models are single-field models~\cite{Maciejewski:2008hj,Bars:2011aa,Bars:2011mh,Bars:2012mt,Kamenshchik:2013dga,Fre:2013vza,Boisseau:2015hqa,Kamenshchik:2015cla,Afanasev:2025cpi} and $F(R)$ gravity models~\cite{Paliathanasis:2016tch,Carloni:2004kp,Faraoni:2017ecj,Banijamali:2019gry,Paliathanasis:2025sft,Shubina:2025ocl}. Some two-field integrable cosmological model have been proposed in Refs.~\cite{Paliathanasis:2018vru,Chimento:1998ju,Paliathanasis:2014yfa,Giacomini:2021xsx,Christodoulidis:2021vye,Ivanov:2021ily,Russo:2023nir,Ivanov:2024nnd}.
The integrability of many cosmological models with minimally coupled scalar fields has been proved by solving evolution equations using a suitable parametrization of time~\cite{Fre:2013vza}.
The integrability of the Starobinsky model as well as the integrability of $R+R^n$ and pure $R^n$ cosmological models has been investigated by the singularity analysis~\cite{Paliathanasis:2016tch}.

The interrelation between the Jordan and Einstein frames is useful for constructing integrable models with nonminimally coupled scalar fields~\cite{Kamenshchik:2013dga}. The inverse method that uses the integrability of cosmological models with nonminimal coupling to prove the integrability of the corresponding models in the Einstein frame has been used as well~\cite{Bars:2011aa,Bars:2011mh,Bars:2012mt,Boisseau:2015hqa,Kamenshchik:2015cla,Ivanov:2024yjr,Ivanov:2024nnd}.

The integrability of pure $R^2$ models with a minimally coupled scalar or phantom scalar field without a potential has been proven for the Friedmann-Lemaitre-Robertson-Walker (FLRW) and Bianchi I metrics~\cite{Ivanov:2021ily,Ivanov:2023uvh}.
In this paper, we investigate the pure $R^2$ model with radiation and find the general solution of the trace equation in the spatially flat FLRW metric. We analyze how the evolution of the Hubble parameter depends on the sign of the radiation energy density.

 Radiation can be represented as a nonminimally coupled scalar field, whose action includes the induced gravity term and the fourth-order potential. We demonstrate the integrability of this $F(R,\chi)$ model in the spatially flat FLRW metric. We also construct the corresponding two-field model in the Einstein frame, thereby finding a new integrable CCM.

\section{$R^2$ model with radiation}

We start with the action
\begin{equation}
\label{initial_actionR2}
S = \int d^4 x \sqrt{-g} \left(F_0 R^2 + \mathcal{L}_{rad}\right),
\end{equation}
where $F_0$ is a positive dimensionless constant, the Lagrangian $\mathcal{L}_{rad}$ represents radiation.

Variation of action (\ref{initial_actionR2}) with respect to the metric $g^{\mu \nu}$ gives us the following system of equations:
\begin{equation}
\label{initial_equations}
2 F_0 R \left(R_{\mu \nu} - \frac14 g_{\mu \nu} R\right) - 2 F_0 \left(\nabla_\mu \nabla_\nu - g_{\mu \nu} \Box \right) R = \frac12 T_{\mu \nu},
\end{equation}
where $\nabla_\nu$ is the covariant derivative, $\Box = g^{\mu \nu} \nabla_\mu \nabla_\nu$,  and
\begin{equation}
T_{\mu \nu} = 2 \frac{\partial \mathcal{L}_{rad}}{\partial g^{\mu \nu}} - g_{\mu \nu} \mathcal{L}_{rad}.
\end{equation}

The stress-energy tensor $T_{\mu \nu}$ of radiation is traceless:
\begin{equation}
\label{tracerad}
g^{\mu \nu} T_{\mu \nu} = 0.
\end{equation}

From Eqs.~(\ref{initial_equations}) and (\ref{tracerad}), we obtain the trace equation:
\begin{equation}
\label{trace_equation}
\Box R = 0.
\end{equation}

Also, the standard conservation law follows from Eq.~(\ref{initial_equations}):
\begin{equation}
\label{conservation_law}
\nabla_\mu T^{\mu \nu} = 0.
\end{equation}

\section{Equations in the FLRW metric}

We use the spatially flat FLRW metric, with its line element being
\begin{equation}
\label{FLRW}
ds^2 ={} -dt^2 + a^2(t)\left(dx^2 + dy^2 + dz^2\right),
\end{equation}
where $a(t)$ is the scale factor, which is positive by definition.

With such a metric, we write the stress-energy tensor as a perfect radiation-like fluid at rest:
\begin{equation}
T_{\mu \nu} = \rho(t)\,\mathrm{diag}\left(1, \frac13 a^2(t), \frac13 a^2(t), \frac13 a^2(t)\right),
\end{equation}
where $\rho(t)$ is the energy density. To explore the solution space more fully we do not require that $\rho(t)$ should be positive.

Using the relations
\begin{equation*}
R_{00}= {}-3\left(\dot{H} + H^2\right),\qquad
R_{11}= a^2\left(\dot{H} + 3 H^2\right),
\end{equation*}
\begin{equation}
\label{Ra}
R = 6\left(\dot{H} + 2 H^2\right)=6 \left(\frac{\ddot{a}}{a} + \left[\frac{\dot{a}}{a}\right]^2\right),
\end{equation}
where ``dots'' denote derivatives with respect to time $t$, and $H = \dot{a} / a$ is the Hubble parameter,
we present Eq.~(\ref{initial_equations}) as the following system of equations:
\begin{equation}
\label{flrw_00}
18 F_0 \left(2 H \ddot{H} - \dot{H}^2 + 6 H^2 \dot{H}\right) = \frac12 \rho,
\end{equation}
\begin{equation}
\label{flrw_11}
6 F_0 \left(2 \dddot{H} + 12 H \ddot{H} + 9 \dot{H}^2 + 18 H^2 \dot{H}\right) = -\frac16 \rho.
\end{equation}

For future use, we rewrite Eq.~(\ref{flrw_00}) as follows:
\begin{equation}
\label{flrw_00_r}
12 H \dot{R} - R^2 + 12 H^2 R=12 H \dot{R} - 6\dot{H} R = \frac{1}{F_0} \rho.
\end{equation}

The conservation law~(\ref{conservation_law}) in the FLRW metric takes the form
\begin{equation}
\label{equrho}
\dot{\rho} + 4 H \rho = 0.
\end{equation}
It is immediately integrable:
\begin{equation}
\label{rho_soln}
\rho(t) = \rho_0 \left(\frac{a_0}{a(t)}\right)^4,
\end{equation}
where $\rho_0 = \rho(t_0)$ and $a_0 = a(t_0)$ and $t_0$ is some moment of time. Note that $\rho(t)\neq 0$ at any $t$ if $\rho_0\neq 0$.

Equations (\ref{Ra}), (\ref{flrw_00_r}), and (\ref{equrho}) define a system of three first-order differential equations in $H$, $R$, and $\rho$.

The trace equation~(\ref{trace_equation}) in the FLRW metric can be written as the following third-order differential equation in $H$:
\begin{equation}
\label{TraceR_FLRW}
\ddot{R} + 3 H \dot{R} = 0, \quad\Leftrightarrow \quad \dddot{H}={}-7H\ddot{H}-12H^2\dot{H}-4H{\dot{H}}^2.
\end{equation}

Integrating this equation, we connect functions $\dot{R}(t)$ and $a(t)$:
\begin{equation}
\label{dot_r_eq}
\dot{R}(t) = \dot{R}_0 \left(\frac{a_0}{a(t)}\right)^3,
\end{equation}
where $\dot{R}_0 = \dot{R}(t_0)$.

This result has an important consequence: for a physically sensible $a(t)$ (i.e. $a > 0$), the function $\dot{R}(t)$ never changes its sign, and it is equal to zero only if
$\dot{R}_0 = 0$. This implies that the function $R(t)$ is either strictly monotonic or constant.

In what follows, we will consider only the case of $R>0$. Zero and negative $R$ are not suitable for the description of pre-inflation and generically lead to instabilities~\cite{Ivanov:2023uvh}. Moreover, our model is supposed to be an approximation to a more realistic model (such as the Starobinsky inflation model) in the large $R$ regime. $R$ approaching zero in a finite time would make our model an inadequate approximation.

\section{Solutions with a constant $R$}
In the case of $R = R_0 = \mathrm{const}$, the relation (\ref{Ra}) gives the following equation:
\begin{equation}
\label{eqH}
\dot{H} + 2 H^2 = \frac{R_0}{6}.
\end{equation}

In dependence on the sign of~$\dot{H}$, solutions of Eq.~(\ref{eqH}) can be written as follows:
\begin{itemize}
\item $\dot{H} > 0$:
\begin{equation}
H(t) = \sqrt{\frac{R_0}{12}} \tanh \left(\sqrt{\frac{R_0}{3}}(t - t_0)\right),
\end{equation}
\begin{equation}
a(t) = a_1 \sqrt{\cosh\left(\sqrt{\frac{R_0}{3}}(t - t_0)\right)};
\end{equation}
\item $\dot{H} = 0$:
\begin{equation}
H(t) = \pm \sqrt{\frac{R_0}{12}},
\end{equation}
\begin{equation}
a(t) = a_0 e^{\pm \sqrt{\frac{R_0}{12}}(t-t_0)};
\end{equation}
\item $\dot{H} <  0$:
\begin{equation}
H(t) = \sqrt{\frac{R_0}{12}} \coth \left(\sqrt{\frac{R_0}{3}}(t - t_0)\right),
\end{equation}
\begin{equation}
a(t) = a_1 \sqrt{\sinh \left(\sqrt{\frac{R_0}{3}}(t - t_0)\right)}.
\end{equation}
\end{itemize}
Here $a_1$ is a constant.

Using these explicit expressions of $a(t)$ and Eq.~(\ref{rho_soln}), we obtain $\rho(t)$ explicitly. The same expressions can be obtained
as solutions of Eq.~(\ref{flrw_00_r}) with a constant $R = R_0$:
\begin{equation}
\label{constraint_const_r}
\rho = {}-6 R_0 F_0\dot{H} \,.
\end{equation}

This equation connects the signs of $\dot{H}$ and $\rho$ and shows that the ``$\tanh$'' solution corresponds to a negative energy density of radiation, the ``$\coth$'' solution corresponds to a positive energy density, and a constant $H$ corresponds to the case without radiation.

\section{Solutions with non-constant $R$}

If the Ricci scalar isn't a constant, then it is a monotonic function of time. So, we can consider $a$ as a function of $R$ and get
\begin{equation*}
\begin{aligned}
\dot{a} &= \dot{R} \frac{d a}{d R} = \dot{R}_0 \left(\frac{a_0}{a}\right)^3 \frac{d a}{d R},\\
\ddot{a} & = \dot{R}^2 \frac{d^2 a}{d R^2} + \ddot{R} \frac{d a}{d R} = \frac{\dot{R}_0^2 a_0^6}{a^7} \left(a \frac{d^2 a}{d R^2} - 3\left(\frac{d a}{d R}\right)^2\right).
\end{aligned}
\end{equation*}
Here we have used Eq.~(\ref{dot_r_eq}) to express $\dot{R}$ in terms of $a$. Now, using relation (\ref{Ra}),
we obtain
\begin{equation*}
\begin{aligned}
R &= 6 \frac{\dot{R}_0^2 a_0^6}{a^8} \left(a \frac{d^2 a}{d R^2} - 2\left(\frac{d a}{d R}\right)^2\right) = 6 \frac{\dot{R}_0^2 a_0^6}{a^5} \frac{d}{d R} \left(\frac{1}{a^2} \frac{d a}{d R}\right) \\
  &= {}-6 \dot{R}_0^2 \left(\frac{a_0}{a}\right)^5 \frac{d^2}{d R^2} \left(\frac{a_0}{a}\right).
\end{aligned}
\end{equation*}

After some algebraic manipulations, one obtains
\begin{equation}
\label{EqR}
\left(\frac{a_0}{a}\right)^5 \frac{d^2}{d R^2} \left(\frac{a_0}{a}\right) = {}-\frac{R}{6 \dot{R}_0^2}\,.
\end{equation}

Let us now introduce a new independent variable $x$, defined as
\begin{equation}
\label{x_def}
x = \frac12 \ln \frac{R}{R_0}.
\end{equation}
From $R_0 = R(t_0)$, it follows $x(t_0) = 0$. In terms of $x$, Eq.~(\ref{EqR}) reads:
\begin{equation}
\label{equvarthetax}
\left(e^{-x} \vartheta\right)^5 \left(e^{-x} \vartheta_{xx}'' - 2 e^{-x} \vartheta_{x}'\right) = -\frac{2 R_0^3}{3 \dot{R}_0^2},
\end{equation}
where $\vartheta(R) = a_0 / a$, \ $\vartheta_{x}'=\frac{d\vartheta}{dx}$.

To get a second-order autonomous differential equation we introduce a new dependent variable
\begin{equation}
\label{w_def}
w = w_0 \vartheta e^{-x} = w_0 \frac{a_0}{a} e^{-x},
\end{equation}
with
\begin{equation}
\label{w0}
w(t_0) = w_0 = \left(\frac{3 \dot{R}_0^2}{2 R_0^3}\right)^{1/6}.
\end{equation}

In terms of $w$, we get Eq.~(\ref{equvarthetax}) as
\begin{equation}
\label{equw}
w^5 \left(w_{xx}'' - w\right) + 1 = 0.
\end{equation}

We integrate Eq.~(\ref{equw}) once with respect to $x$ and get
\begin{equation}
\label{energy_condition}
\frac12 \left(w_{x}'\right)^2 - \frac{w^2}{2} - \frac{1}{4 w^4} = E,
\end{equation}
where $E$ is an integration constant.

This is, essentially, the energy conservation law for a fictitious particle with coordinate $w$ and velocity $w_{x}'$.
Equation (\ref{energy_condition}) gives
\begin{equation}
\label{dwdx}
w_{x}'= \sigma_w \sqrt{w^2 + \frac{1}{2 w^4} + 2 E},
\end{equation}
where $\sigma_w = \pm 1$.

So, we have shown that the dynamics of our model is reduced to the one-dimensional dynamics of a particle in the potential
\begin{equation}
W(w) = -\frac{w^2}{2} - \frac{1}{4 w^4}.
\end{equation}
This potential is negative everywhere, and has a single extremum (maximum) at $w=1$: $W_{max} = W(1) = -3/4$.

Let us now recover the initial dynamical quantities. From Eq.~(\ref{w_def}), one obtains
\begin{equation}
\label{a_sol_var_r}
a = a_0 \frac{w_0}{w} e^{-x},
\end{equation}
and it follows from Eq.~(\ref{x_def}):
\begin{equation}
\label{r_sol_var_r}
R = R_0 e^{2 x}.
\end{equation}
Using Eqs.~(\ref{dot_r_eq}) and~(\ref{a_sol_var_r}), we also obtain
\begin{equation}
\label{dot_r_sol_var_r}
\dot{R} = \dot{R}_0 \left(\frac{w}{w_0}\right)^3 e^{3 x}.
\end{equation}

Equations~(\ref{r_sol_var_r}) and~(\ref{dot_r_sol_var_r}) allow us to write $x$ and $w$ as functions of $t$. Taking the derivative of Eq.~(\ref{r_sol_var_r}) with respect to $t$ and comparing it with Eq.~(\ref{dot_r_sol_var_r}), one gets
\begin{equation*}
2 R_0 e^{2x} \dot{x} = \dot{R}_0 \left(\frac{w}{w_0}\right)^3 e^{3 x},
\end{equation*}
which gives us the following first-order ODE:
\begin{equation}
\label{dot_x}
\dot{x} = \frac{\dot{R}_0}{2 R_0} \left(\frac{w}{w_0}\right)^3 e^x = \mathrm{sgn}\left(\dot{R}_0\right) \sqrt{\frac{R_0}{6}} w^3 e^{x}.
\end{equation}
An analogous formula for $\dot{w}$ immediately follows from Eqs.~(\ref{dwdx}) and (\ref{dot_x}),
\begin{equation}
\label{dot_w}
\dot{w} = w_{x}' \dot{x} = \sigma_w \, \mathrm{sgn}\left(\dot{R}_0\right) \sqrt{\frac{R_0}{6}} w^3 e^{x} \sqrt{w^2 + \frac{1}{2 w^4} + 2 E}.
\end{equation}

Finally, we write the formula for the Hubble parameter in terms of $x$ and $w$. Using Eqs.~(\ref{a_sol_var_r}), (\ref{dot_x}), and (\ref{dot_w}), one obtains
\begin{equation}
\label{H_sol}
H = {} -\frac{\dot{w}}{w} - \dot{x} ={} -\mathrm{sgn}\left(\dot{R}_0\right) \sqrt{\frac{R_0}{6}} w^2 e^{x}\left(\sigma_w \sqrt{w^2 + \frac{1}{2 w^4} + 2 E} + w \right).
\end{equation}

Let us also connect the constant $E$ with initial conditions. We can do that by inserting the obtained solutions for $a$, $H$, and $R$ into the constraint equation~(\ref{flrw_00_r}). After some algebraic manipulations, one can finally obtain the relation:
\begin{equation}
E = \frac{\rho_0}{6 F_0} \sqrt[3]{\frac{3}{2 \left(\dot{R}_0\right)^4}}.
\end{equation}

So, the pure $R^2$ gravity model with traceless matter is integrable in the spatially flat FLRW metric.

\section{Analysis of the solutions}
\subsection{Solutions without radiation}
If $\rho_0=0$, then Eq.~(\ref{TraceR_FLRW}) is a consequence of the second-order equation Eq.~(\ref{flrw_00}). In this case, the general solution can be found explicitly as the inverse function $t(H)$:
\begin{equation}\label{ht_rho0}
\begin{split}
    t-t_0&=\frac{{2}^{2/3}}{12{C}^{2/3}} \left[ 2\,\sqrt {3}\arctan \left(
{\frac { 2^{4/3}\sqrt{H}+\sqrt[3]{C} }{\sqrt{3}\sqrt[3]{C}}} \right) \right.\\& {} +2\,\sqrt {3}\arctan \left(
\,{\frac { {2}^{4/3}\sqrt {H}-\sqrt [3]{C}}{\sqrt {3}\sqrt [3]{C}}} \right) -2\,\sqrt {3}\arctan \left( {\frac{ {2}^{5/3}H+{C}^{2/3}}{\sqrt{3}{C}^{
2/3}}} \right) \\&{} -2\,\ln  \left( -{2}^{2/3}\sqrt [3]{C}+2\,\sqrt {H} \right) +\ln  \left(\sqrt
[3]{2}{C}^{2/3}+{2}^{2/3}\sqrt [3]{C}\sqrt {H}+2H \right) \\&{}+2\,\ln  \left( {2}^{2/3}\sqrt [3]
{C}+2\,\sqrt{H} \right) -\ln  \left(\sqrt [3]{2}{C}^{2/3} -{2}^{2/3}\sqrt [3]{C}\sqrt {H}+2\,H \right) \\&\left.{}-2\,\ln \left( -\sqrt [3]{2}{C}^{2/3}+2\,H \right) +\ln  \left( {2}^{2/3}{C}^{4/3}+2^{4/3}{C}^{2/3}H+4{H}^{2} \right)  \right],
\end{split}
\end{equation}
where $t_0$ and $C$ are integration constants.

\subsection{Asymptotic behavior of $x$ and $w$}

In the case of a nonzero $\rho_0$, the analysis of solutions is more complicated.

In the large $w$ limit, Eq.~(\ref{dwdx})
\begin{equation*}
w_{x}'= \sigma_w w + O\left(\frac{1}{w}\right),
\end{equation*}
can be integrated and gives
\begin{equation*}
\sigma_w \ln \frac{w}{w_0} \approx x.
\end{equation*}

To recover the $w(t)$ dependency, we use Eq.~(\ref{dot_w}) in the same large $w$ limit:
\begin{equation*}
\dot{w} \approx \sigma_w \mathrm{sgn}\left(\dot{R}_0\right) \sqrt{\frac{R_0}{6}} w^4 \left(\frac{w}{w_0}\right)^{\sigma_w}.
\end{equation*}
Integration with respect to $x$ leads to
\begin{equation*}
\left(\frac{w_0}{w}\right)^{3 + \sigma_w} \approx 1 - \sigma_w \mathrm{sgn}\left(\dot{R}_0\right) (3 + \sigma_w)\sqrt{\frac{R_0}{6}} w_0^3 (t - t_0),
\end{equation*}
from which it follows
\begin{equation*}
w(t) \approx w_0 \left(1 - \sigma_w \mathrm{sgn}\left(\dot{R}_0\right) (3 + \sigma_w)\sqrt{\frac{R_0}{6}} w_0^3 (t - t_0)\right)^{-\frac{1}{3 + \sigma_w}}.
\end{equation*}
We see that for any choice of $\sigma_w$, the function $w(t)$ approaches infinity in finite time (either in the future or in the past).

Let us substitute our large-$w$-limit $x(w)$ dependency into Eq.~(\ref{r_sol_var_r}):
\begin{equation}
R \approx R_0 \left(\frac{w}{w_0}\right)^{2 \sigma_w} \approx \left(1 - \sigma_w \mathrm{sgn}\left(\dot{R}_0\right) (3 + \sigma_w)\sqrt{\frac{R_0}{6}} w_0^3 (t - t_0)\right)^{-\frac{2 \sigma_w}{3 + \sigma_w}}.
\end{equation}
The consequences of this result are substantial: we see that if $\sigma_w = -1$, then $R$ tends to zero as $w$ tends to infinity (which happens, as we have just shown, in finite time). Therefore, the branch $\sigma_w = -1$ corresponds to $R$ changing its sign. As we have discussed before, this kind of behavior is undesirable, because it leads to instabilities and necessitates the inclusion of the Einstein-Hilbert term, since the system leaves the large $R$ regime in finite time. The only situation where $\sigma_w = -1$ can be allowed is the one in which $w$ never tends to infinity during its evolution. As we will see in what follows, this situation can be realized for certain initial conditions. If $\sigma_w = 1$, however, we get a singular behavior: $R$ decreases from infinity in finite time or goes to infinity in finite time. This kind of behavior is allowed. It may simply correspond to the initial singularity (or the ``big crunch'' singularity).

Now we consider the case of $w \rightarrow 0$. It follows from Eq.~(\ref{dwdx}) that
\begin{equation*}
w_{x}' = \sigma_w \frac{1}{\sqrt{2} w^2} + O(w^4).
\end{equation*}
Integrating it, we obtain
\begin{equation*}
\sigma_w \frac{\sqrt{2}}{3} \left(w^3 - w_0^3\right) \approx x \implies e^{x} \approx e^{\sigma_w \frac{\sqrt{2}}{3} \left(w^3 - w_0^3\right)}.
\end{equation*}
We see that, at the lowest order in $w$, $e^{x}$ is just some constant very close to $1$. So, after plugging our results into Eq.~(\ref{dot_w}), we obtain
\begin{equation*}
\dot{w} \approx \sigma_w \mathrm{sgn}\left(\dot{R}_0\right) \sqrt{\frac{R_0}{12}} w,
\end{equation*}
from which we get
\begin{equation*}
w(t) \approx w_0 \exp\left(\sigma_w \mathrm{sgn}\left(\dot{R}_0\right) \sqrt{\frac{R_0}{12}}(t-t_0)\right).
\end{equation*}
We see that $w$ tends to zero exponentially fast (of course, if $\dot{w} < 0$), but never reaches it. If $\dot{w} > 0$, then $w$ leaves the neighborhood of zero exponentially fast.

Plugging our small-$w$-limit $x(w)$ dependency into Eq.~(\ref{r_sol_var_r}), we obtain
\begin{equation*}
R \approx R_0.
\end{equation*}
So, the ``$w \rightarrow 0$'' limit corresponds to constant curvature, or, in other words, to de Sitter limit. Note that this statement is true irrespective of the choice of $\sigma_w$.

\subsection{Dynamics of $w$}
Let us investigate the equation for $w_{x}'$. Rewriting Eq.~(\ref{dwdx}) a bit, we get
\begin{equation*}
w_{x}' = \sigma_w \sqrt{2\left(E - W(w)\right)}.
\end{equation*}
As it is known from classical mechanics, roots of the equation $E - W = 0$ correspond to the turning points, at which $w_{x}'$ goes through zero, and $\sigma_w$ changes sign. Let us find those roots.

So, we have the equation
\begin{equation*}
E - W(w) = E + \frac{w^2}{2} + \frac{1}{4 w^4} = 0.
\end{equation*}
Since we know that the potential is no larger than $-3/4$, we conclude that for $E > -3/4$ there exist no roots of this equation.

To study the case of $E \leq -3/4$, we multiply the $E - W = 0$ equation by $4/w^2$ and obtain
\begin{equation*}
\frac{1}{w^6} + 4 E \frac{1}{w^2} + 2 = 0.
\end{equation*}
This is a depressed cubic equation with respect to the variable $y = 1/w^2$. If $E \leq -3/4$, the positive roots of the equation are given by the formula
\begin{equation}
\label{turning_points}
y_k = \frac{1}{r_k^2} = 2 \sqrt{\frac{4\left|E\right|}{3}} \cos \left(\frac13 \arccos \left[-\left(\frac{3}{4 \left|E\right|}\right)^{3/2}\right] - \frac{2\pi k}{3}\right), \quad k = 0, 1.
\end{equation}
Here we denote roots of the $E - W = 0$ equation in terms of $w$ as $r_k$.
As $\left|E\right| \rightarrow \infty$, $y_0 \rightarrow \infty$ and $y_1 \rightarrow 0$. This implies that $y_0$ corresponds to the smaller root in terms of $w$, while $y_1$ corresponds to the larger one.

\begin{figure}[h!]
\centering
\includegraphics[width=1\linewidth]{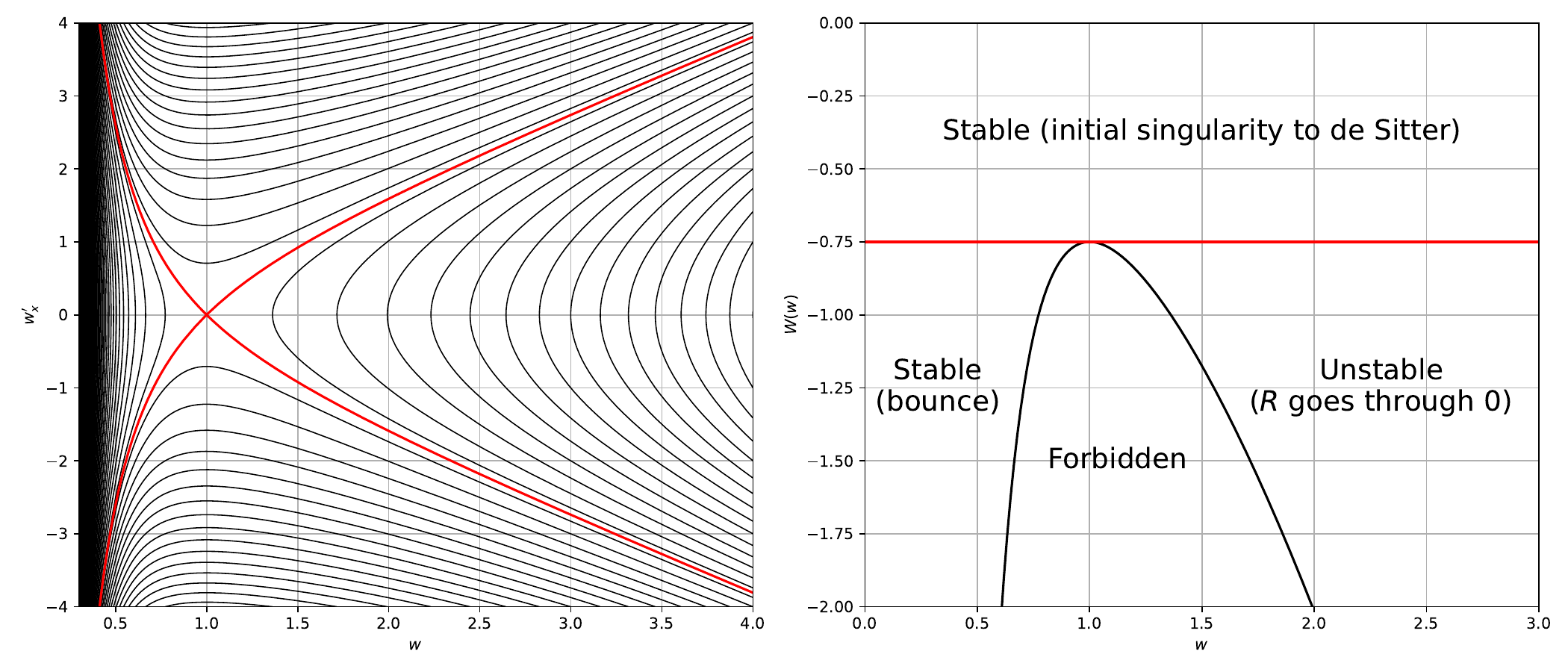}
\caption{$w_{x}'$-$w$ phase diagram (left) and the potential $W(w)$ in black (right). On the left plot, each continuous black line corresponds to a particular value of $E$. On the right plot, for the uppermost region $\dot{w}_0$ is assumed to be negative and $\sigma_w = +1$. On both plots, red lines correspond to the separatrix ($E = -3/4$).\label{fig1}}
\end{figure}

So, the dynamics of $w$ falls into one of four cases (see also Fig.~\ref{fig1}):
\begin{enumerate}
\item $E > -\frac34$. This is the case of unbounded movement.
\item $E = -\frac34$. This is the separatrix case.
\item $E < -\frac34$, $w_0 < r_0$. This is the stable case with reflection of $w$.
\item $E < -\frac34$, $w_0 > r_1$. This is the unstable case with reflection of $w$, with $R$ changing its sign.
\end{enumerate}

\subsection{Behavior of the Hubble parameter $H$}
Now we turn our attention to Eq.~(\ref{H_sol}). If we are interested in describing (eventually) expanding Universe, then we must understand, which initial conditions provide us with positive $H$.
Let us connect initial values of the Hubble parameter and its derivatives with $E$, $R_0$ and $\dot{R}_0$. It follows from Eqs.~(\ref{H_sol})  that
\begin{equation}\label{H0}
H_0 = {} -\mathrm{sgn}\left(\dot{R}_0\right) \sqrt{\frac{R_0}{6}}w_0^2\left(\sigma_w \sqrt{w_0^2 + \frac{1}{2 w_0^4} + 2 E} + w_0 \right).
\end{equation}

Relation (\ref{w0}) connects $w_0$ with $R_0$ and $\dot{R}_0$.
Using Eq.~(\ref{Ra}), we get
\begin{equation}
\dot{H}_0=\frac{R_0}{6}-2H_0^2, \qquad \ddot{H}_0=\frac{\dot{R}_0}{6}-4H_0\dot{H}_0\,.
\end{equation}

\begin{figure}[h!]
\centering
\includegraphics[width=1\linewidth]{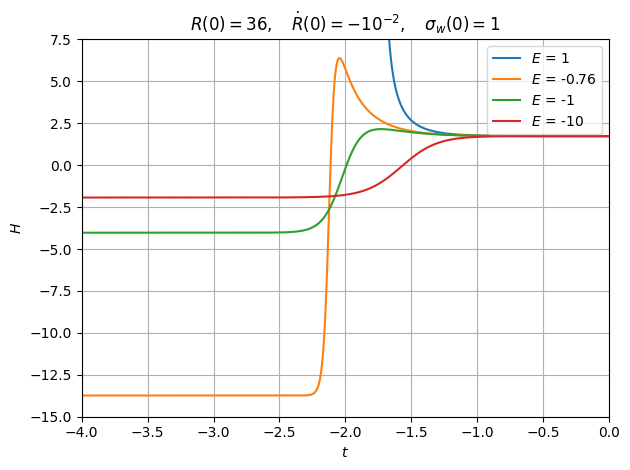}
\caption{The behavior of the Hubble parameter $H(t)$ at $\sigma_w=1$, $R_0=36$, $\dot{R}_0=-10^{-2}$, and different values of $E$.
\label{figHt}}
\end{figure}

So, to get the initial condition of Eq.~(\ref{TraceR_FLRW}) one should fix not only parameters $E$, $R_0$ and $\dot{R}_0$, but also $\sigma_w$. Numerical solutions of Eq.~(\ref{TraceR_FLRW}) are presented in Fig.~\ref{figHt}.

Let us describe the behavior of $H$ in detail.
Firstly, we consider the case $E > -3/4$, in which there are no turning points of $w$. So, to get the ``initial singularity to de Sitter'' behavior (we remind that it means that $w$ goes from infinity to zero with $\sigma_w = 1$), it is obvious to require that $\dot{w}$ is negative. Therefore, from Eq.~(\ref{dot_w}) it follows that $\dot{R}_0$ must be negative. Substituting this choice of signs of $\dot{R}_0$ and $\sigma_w$ into Eq.~(\ref{H_sol}), we get
\begin{equation*}
H = \sqrt{\frac{R_0}{6}} w^2 e^{x} \left(\sqrt{w^2 + \frac{1}{2 w^4} + 2 E} + w\right).
\end{equation*}
We see that $H>0$ for any $w > 0$ and any $x$, as expected.

At $E = -\frac34$, one of particular solutions is
\begin{equation}\label{H1t}
   H=\frac{1}{t-t_0}\,.
\end{equation}
The corresponding Ricci scalar is
\begin{equation}
R=\frac{6}{\left(t-t_0\right)^2}\,.
\end{equation}
This solution has a singular point at $t=t_0$ and tends to the Minkowski space-time at large $t$.
This is the separatrix case with $w = \mathrm{const} = 1$. It corresponds to the maximum of the potential $W$. It is quite obvious that this solution is (dynamically) unstable.

As we've discussed before, in the case of $E < -3/4$ the turning points of $w$ occur, given by Eq.~(\ref{turning_points}). Again, we emphasize that the region $w \geqslant r_1$ is fundamentally physically unstable, so we will consider only the case of $w \leqslant r_0$. In this region, $\sigma_w$ inevitably changes, so we must consider both situations. For the case of $\sigma_w = 1$, we get
\begin{equation*}
H = {}-\mathrm{sgn}\left(\dot{R}_0\right)\sqrt{\frac{R_0}{6}} w^2 e^{x} \left(\sqrt{w^2 + \frac{1}{2 w^4} + 2 E} + w\right).
\end{equation*}

Obviously, the sign of $H$ remains unchanged in this case. On the other hand, in the case of $\sigma_w = -1$, we have
\begin{equation*}
H = {}-\mathrm{sgn}\left(\dot{R}_0\right)\sqrt{\frac{R_0}{6}} w^2 e^{x} \left(-\sqrt{w^2 + \frac{1}{2 w^4} + 2 E} + w\right).
\end{equation*}
Now it is possible that the expression inside the brackets equals zero for some value of $w$. It is easy to show that this value is
\begin{equation}
w_{bounce} = \left(-4 E\right)^{-1/4} < r_0.
\end{equation}

Therefore, at $w = w_{bounce}$ (if $\sigma_w = -1$) the sign of $H$ changes. Thus, we conclude that initial conditions corresponding to $E < -3/4$ and $w_0 < r_0$ are related to the ``bouncing'' solutions. To get a reasonable ``negative-$H$-to-positive-$H$'' behavior, we can actually take any choice of signs of $\dot{R}_0$ and $\sigma_w$. That is because the direction of movement of $w$ depends on the product $\sigma_w \mathrm{sgn}\left(\dot{R}_0\right)$ (see Eq.~(\ref{dot_w})). If we take this product to be $+1$, we get the ``left-to-right'' evolution of $w$, and so the limit $w \rightarrow 0$ will correspond to the past and, according to the formulae we've just obtained, to negative $H$. On the contrary, taking this product to be $-1$, we get the ``right-to-left'' evolution of $w$, and so the limit $w \rightarrow 0$ corresponds to the future and to positive $H$. Obviously, both of these choices belong to a single curve in the phase space, we simply consider different initial times when we make any particular choice of the signs.
In Fig.~\ref{figHt}, we demonstrate how the behavior of the Hubble parameter depends on $E$.

\section{A non-minimally coupled scalar field as radiation}

Up to now, we do not specify the Lagrangian of radiation. In this section, we choose it in the following form:
 \begin{equation}
\mathcal{L}_{rad}={}-\frac{b\chi^2}{12}R-\frac b2{g}^{\mu\nu}\partial_{\mu}\chi\partial_{\nu}\chi-c\chi^4,
\label{Lrad}
\end{equation}
where $b=\pm 1$, $c$ is a constant, and $\chi$ is a scalar field.

Action (\ref{initial_actionR2}) can be rewritten as the following action of two-field model:
\begin{equation}
\label{action2f}
    S=\int d^4x\sqrt{-g}\left(UR-\frac b2{g}^{\mu\nu}\partial_{\mu}\chi\partial_{\nu}\chi-V\right),
\end{equation}
where
\begin{equation}
\label{Uchi}
U=2F_0\sigma-\frac{b}{12}\chi^2,
\end{equation}
\begin{equation}
\label{Vchi}
V=F_0\sigma^2+c\chi^4.
\end{equation}

Evolution equations have the following form:
\begin{equation}\label{Equ2f}
U\left(R_{\mu \nu}- \frac R2 g_{\mu \nu} \right) = \nabla_\mu \nabla_\nu U - g_{\mu \nu} \Box U +  \frac b2 \partial_\mu\chi\partial_\nu\chi
-\frac{g_{\mu \nu}b}{4}g^{\alpha \beta}\partial_\alpha\chi\partial_\beta\chi -\frac{g_{\mu \nu}}{2} V\,.
\end{equation}

The trace equation is
\begin{equation}\label{tracetwofields}
    UR-3\Box U-\frac b2 g^{\alpha \beta}\partial_\alpha\chi\partial_\beta\chi -2V=0.
\end{equation}
The field equations are
$\sigma=R$
and
\begin{equation}\label{equchi}
  b \Box\chi=V_{,\chi}-U_{,\chi}R\,.
\end{equation}

Using Eqs.~(\ref{Uchi}) and (\ref{equchi}), we get
\begin{equation}\label{boxU}
    \Box U=2F_0\Box\sigma-\frac{1}{6}\chi\left(V_{,\chi}-U_{,\chi}R\right)-\frac{b}{6}g^{\alpha \beta}\partial_\alpha\chi\partial_\beta\chi\,.
\end{equation}
Substituting (\ref{Uchi}), (\ref{Vchi}), and (\ref{boxU}) into Eq.~(\ref{tracetwofields}), we obtain Eq.~(\ref{trace_equation}). It proves that the Lagrangian (\ref{Lrad}) describes radiation.

In the spatially flat FLRW metric with the conformal time $\tau$,
\begin{equation*}
ds^2=a^2(\tau)\left(-d\tau^2+dx^2 + dy^2 + dz^2\right),
\end{equation*}
we get
\begin{equation}
\label{equ00}
6h^2U+6h\frac{dU}{d\tau}=\frac{b}{2}\left[\frac{d\chi}{d\tau}\right]^2+Va^2\,,
\end{equation}
\begin{equation}
\label{equ11}
2U\left(2\frac{dh}{d\tau}+h^2\right)+4h\frac{dU}{d\tau}+2\frac{d^2U}{d\tau^2}=-\frac{b}{2}\left[\frac{d\chi}{d\tau}\right]^2+Va^2\,,
\end{equation}
where $h=d\ln(a)/d\tau$.

From Eq.~(\ref{flrw_00}), we obtain
\begin{equation}\label{rho}
    \rho(t) =\frac b2\left(\dot\chi-H\chi\right)^2+c\chi^4=\frac {b}{2a^2}\left(\frac{d\chi}{d\tau}-h\chi\right)^2+c\chi^4.
\end{equation}

The field equations take the following form:
\begin{equation}
\label{equphi}
\frac{d^2\chi}{d\tau^2}+2h\frac{d\chi}{d\tau}+V_{,\chi}a^2-\frac{6}{a^2}\frac{d^2a}{d\tau^2}U_{\chi}=0
\end{equation}
and
\begin{equation}
\sigma=R=\frac{6}{a^3}\frac{d^2a}{d\tau^2}\,.
\end{equation}

If we put $\chi=y/a$ and use Eqs.~(\ref{Uchi}) and (\ref{Vchi}), then we get the following equation
\begin{equation}\label{equy}
    \frac{d^2y}{d\tau^2}={}-a^3V'_{,\chi}={}-4c a^3\chi^3={}-4cy^3.
\end{equation}

This equation has the general solution in terms the Jacobi elliptic functions, so, the model (\ref{action2f}) is integrable.

By the conformal transformation of the metric:
\begin{equation}
\label{gJgE}
\tilde{g}_{\mu\nu}=\frac{2U}{M^2_\mathrm{Pl}}{g}_{\mu\nu}\,,
\end{equation}
we obtain corresponding two-field CCM model in the Einstein frame:
\begin{equation}
\label{FRSE}
S_{E}=\int d^4x\sqrt{-\tilde{g}}\left[\frac{M^2_\mathrm{Pl}}{2}\tilde{R}-\frac{3M^2_\mathrm{Pl}}{4U^2}{\tilde{g}^{\mu\nu}}\partial_\mu U\partial_\nu U
-\frac{M^2_\mathrm{Pl}b}{4U}{\tilde{g}^{\mu\nu}}\partial_\mu\chi\partial_\nu\chi-\frac{M^4_\mathrm{Pl}V}{4U^2}\right].
\end{equation}

So, introducing
\begin{equation}
\label{phisigmaFr}
    \phi=\sqrt{\frac{3}{2}}M_\mathrm{Pl}\ln\left(\frac{2U}{M_\mathrm{Pl}^2}\right)=\sqrt{\frac{3}{2}}M_\mathrm{Pl}\ln\left(\frac{1}{M_\mathrm{Pl}^2}\left[4F_0\sigma-\frac{b}{6}\chi^2\right]\right)\,,
\end{equation}
we obtain
\begin{equation}
\sigma=\frac{M_\mathrm{Pl}}{4F_0}\mathrm{e}^{\sqrt{\frac{2}{3}}\,\frac{\phi}{M_\mathrm{Pl}}}+\frac{b\chi^2}{24F_0}\,,
\end{equation}

Therefore,
 \begin{equation}
\label{SE2}
S_{E}=\int d^4x\sqrt{-\tilde{g}}\left[\frac{M^2_\mathrm{Pl}}{2}\tilde{R}-\frac{\tilde{g}^{\mu\nu}}{2}\left[\partial_{\mu}\phi\partial_{\nu}\phi+yb\partial_{\mu}\chi\partial_{\nu}\chi\right]-V_E(\phi,\chi)\right],
\end{equation}
where
\begin{equation}
\label{yphi}
y=\frac{M^2_\mathrm{Pl}}{2U}=\mathrm{e}^{-\sqrt{\frac{2}{3}}\,\frac{\phi}{M_\mathrm{Pl}}},
\end{equation}
and
\begin{equation}
V_E(\phi,\chi)=y^2V=\mathrm{e}^{-2\sqrt{\frac{2}{3}}\,\frac{\phi}{M_\mathrm{Pl}}}\left[F_0\left[\frac{M_\mathrm{Pl}}{4F_0}\mathrm{e}^{\sqrt{\frac{2}{3}}\,\frac{\phi}{M_\mathrm{Pl}}}
+\frac{\chi^2}{24F_0}\right]^2+c\chi^4\right].
\end{equation}

These expressions are valid for any $U>0$.

In the spatially flat FLRW metric, the general solutions of this CCM can be presented in terms of functions obtained in the Jordan frame, so we get a new integrable CCM. The explicit formulae that connect solutions in the Einstein and Jordan frames are given in our paper~\cite{Ivanov:2024nnd}.

\section{Conclusions}

In this paper, we have demonstrated that the pure $R^2$ model with radiation is integrable. We have found the solution of the equation $\Box R=0$ in quadrature. We have also analyzed the behavior of solutions to this equation, and presented the conditions under which bounce solutions exist.

One interesting example of radiation is the scalar field Lagrangian with the induced gravity term and the fourth-order monomial potential. In this case, we have obtained a new integrable $F(R)$ gravity model with a scalar field, as well as the corresponding integrable chiral cosmological model in the Einstein frame. The proposed model can be considered as an $F(R)$ modification of the general relativity model with radiation, described by $\mathcal{L}_{rad}$ that has been investigated in
Refs.~\cite{Boisseau:2015hqa,Kamenshchik:2015cla}. The use of $R^2$ term instead of $M^2_\mathrm{Pl} R/2$ gives a new degree of freedom. So, we get a two-field CCM instead of a single-field model in the Einstein frame. The obtained CCM is integrable in the spatially flat FLRW metric.

The pure $R^2$ gravity model with a minimally coupled scalar field does not have any ghosts when $R > 0$. If the radiation includes the induced gravity term, the situation changes, as the effective gravitational constant now depends on the scalar field. We plan to analyze this issue in future publications.

The investigation was conducted under the state assignment of Lomonosov Moscow State University. V.R.I. is supported by the ``BASIS'' Foundation grant
no. 22-2-2-6-1.

\end{document}